\RequirePackage{ifpdf}
\documentclass[12pt,letterpaper]{JHEP3}
\usepackage{cite}
\usepackage{graphicx}
\usepackage{subfigure}
\graphicspath{{Figures/}} 

\usepackage{color}

\let\normalcolor\relax 

\title{Critical Phenomena in Non-spherically Symmetric  Scalar Bubble Collapse}

\author{Katy Clough\footnote{katy.clough@kcl.ac.uk}, Eugene A. Lim\footnote{eugene.a.lim@gmail.com} \\
Theoretical Particle Physics and Cosmology Group, Physics Department,
King's College London, Strand, London WC2R 2LS, United Kingdom
}

\abstract
{
We investigate numerically the critical behaviour which occurs in the collapse of both spherically symmetric and asymmetric scalar field bubbles with full general relativity. We use a minimally coupled scalar field subject to a ``double well'' interaction potential, with the bubble wall spanning the barrier between two degenerate minima. We find that the spherically symmetric case exhibits Type 2 critical behaviour with the critical index consistent with a value of $\gamma = 0.37$ as expected. In the asymmetric case, we find that again our results are consistent with a value of $\gamma = 0.37$ for the dominant unstable mode. We do not see strong evidence of echoing in the solutions, which could be due to the fact that the coordinates are not well adapted to the echoing behaviour, or due to being still too far from the critical point to properly observe the critical solution.
} 

\begin{document}
\section{Introduction}

Since their discovery by Choptuik \cite{Choptuik:1992jv}, critical phenomena have been studied in many different contexts \cite{Abrahams:1993wa, Healy:2013xia, Choptuik:2003ac, Hilditch:2013cba, Evans:1994pj, Brady:1997fj, Honda:2001xg, Akbarian:2015oaa} -- for an excellent review see \cite{Gundlach:2007gc}.  Briefly, any one parameter $p$ family of initial configurations of the scalar field will evolve to one of the two final end states -- a black hole or the dispersal of the field to infinity. The transition between these two end states occurs at a value of the parameter $p^*$, at which the critical solution exists. 

In a spherically symmetric collapse, the mass of any black hole that is formed from such a collapse follows the critical relation 
\begin{equation}
M \propto (p - p^*)^{\gamma},
\end{equation}
where the scaling constant $\gamma$ is universal in the sense that it does not depend on the choice of family of initial data. For a massless scalar field in a spherically symmetric collapse, $\gamma$ has been numerically determined to be $0.37$. This index does, however, depend on the type of matter considered.

The other key phenomenon which is observed is that of self-similarity in the solutions, or ``scale-echoing''. Close to the critical point, and in the strong field region, the value of any gauge independent field $\phi$ at a position $x$ and time $T$ exhibits the following scaling relation,
\begin{equation}
\phi (x,t) = \phi(e^{\Delta} x, e^{\Delta} T) , \label{eqn:echo}
\end{equation}
where $\Delta$ is a dimensionless constant with another numerically determined value of 3.44 for a massless scalar field in the spherical case. The time coordinate $T$ here is measured ``backwards", that is, it is the difference between the critical time at which the formation of the black hole occurs and the current time, with intervals of time being those of proper time measured by a central observer. What one sees is therefore that, as the time nears the critical time by a factor of $e^{\Delta}$, the same field profile is seen but on a spatial scale $e^{\Delta}$ smaller. This scale-echoing may be either continuous or discrete, but the factors leading a system to either case are not well understood.

In spherically symmetry, the system of equations can be reduced to a 1+1D system, and hence it is widely studied. Beyond spherically symmetry, there has been some recent progress in studying the phenomenon, for example, \cite{Abrahams:1993wa, Healy:2013xia, Choptuik:2003ac, Hilditch:2013cba}, but progress in making firm conclusions has been slower than expected, due to the extremely high refinements required to study the stages of the collapse, which are magnified three-fold in full 3+1 codes. 

This universality in the critical behaviour can be derived by assuming that the critical case is an intermediate attractor of co-dimension one, which, when perturbed, has exactly one unstable mode \cite{Koike:1995jm}. It is not clear whether this will be true in more complex cases beyond spherical symmetry. Linear perturbations of the spherically symmetric case \cite{PhysRevD.59.064031} do not show additional unstable modes, but numerical studies such as that of Choptuik et al. \cite{Choptuik:2003ac} gave hints of further unstable modes, which may be present in the full non-linear regime. Beyond spherical symmetry, work has generally focussed on axisymmetric vacuum collapses of gravitational waves, massless scalar fields, and radiation fluids as in \cite{Abrahams:1993wa, Choptuik:2003ac, Hilditch:2013cba, Gundlach:1999cw,Baumgarte:2015aza}. More recently, Healy and Laguna \cite{Healy:2013xia} considered a fully asymmetric case in full three dimensional simulations for a massless scalar field with $Y^2_1$ spherical harmonic perturbations on a spherical bubble shaped potential.

Adding a self-interaction potential to a massless scalar field is not expected to change the universality of the solution or the critical solution which is approached, therefore one expects that the constants $\gamma$ and $\Delta$ will be unchanged \cite{Chop_conf}. This is because of the assumed self similarity of the critical solution, which means that the field values, and hence the potential, must remain bounded during the evolution \cite{Gundlach:1996eg}. The behaviour is then dominated by the gradient terms in the Lagrangian as the critical solution is approached. The main cases which have been explored beyond the massless case have been massive fields with an $m^2$ mass term such as \cite{Brady:1997fj}, but Honda and Choptuik \cite{Honda:2001xg} also studied sub-critical oscillons in a $\phi^4$ potential. Few papers have considered a more general potential, which is not surprising since it is expected that the results will be the same as those for the massless case. However, the particular case of bubbles set up in multi minima potentials is of interest to early universe cosmologists, as they are natural consequences of phase transitions, as will be discussed below. 

It has not been clear that the standard BSSN formulation of the Einstein equations, combined with the moving puncture gauge commonly used for black hole evolutions, \cite{Campanelli:2005dd, Baker:2005vv}, would be well adapted to the study of this problem. It can be challenging to find a stable choice of gauge parameters near the critical point, and, moreover, it is not clear that any chosen gauge will be ``symmetry seeking'' \cite{Garfinkle:1999cm}, that is, that it will be adapted to observing the scale-echoing phenomena. Recent work by Healy et al \cite{Healy:2013xia} and Akbarian et al. \cite{Akbarian:2015oaa} have indicated that these standard choices may indeed be well adapted for the study of critical phenomena, although the latter paper used a special choice of coordinate grid and made some amendments to the standard puncture gauge. 

\begin{figure}
\begin{center}
\includegraphics[width=15cm]{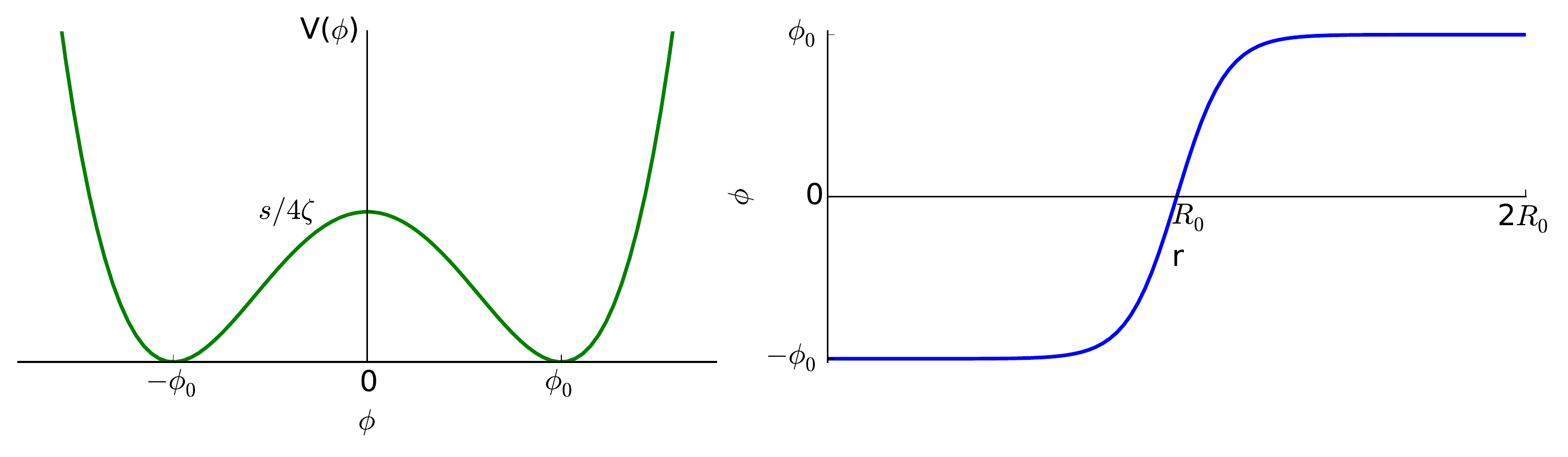}
\caption{The double well potential considered in this paper, and the initial bubble wall profile. 
\label{fig-potential}}
\end{center}
\end{figure}

In this paper, we study the critical collapse and formation of black holes of ``Bubbles'', in both spherically symmetric and perturbed asymmetric cases, in 3+1D using cartesian coordinates and a slightly modified version of the puncture gauge. Bubbles are regions of space bounded by a scalar field domain wall. The domain wall interpolates between the two minima.  We consider the case of a minimally coupled scalar field, subject to the following potential, as shown in Figure \ref{fig-potential}, with two degenerate minima:
\begin{equation}
V(\phi) = \frac{s}{4\zeta \phi_0^4} (\phi^2-\phi_0^2)^2 \label{eqn:potential}
\end{equation}

The potential barrier between the two minima generates a tension $\sigma$ on the bubble wall 
\begin{equation}
\sigma \approx \int_{-\phi_0}^{\phi_0}dx~ \sqrt{2V}.
\end{equation}
The presence of this tension is an important difference between that of a simple ``bubble-like configuration'', say a top-hat, of a scalar-field (either massive or massless) as the field configuration seeks to maintain this tension even as the bubble collapses. Indeed, the tension increases as the bubble becomes smaller, causing the wall's gradient to rapidly increase. Even though the vacuum is degenerate, the bubble collapses due to the pull of both gravity \emph{and} this tension. In other words, without gravity a massless ``bubble'' will disperse while bubbles with tension will coherently collapse.

Our work is motivated by the desire to understand transitions in scalar fields which are subject to multi minima potentials, which are found on cosmological scales in both the early and late universe. 

In the late universe,  cosmological axion fields \cite{Marsh:2015xka} are candidates for Dark Matter. Cosmological axions (as opposed to the QCD axion) are pseudo-bosons which can be described by a real scalar field. If the axion decay constant $f<H_i$ where $H_i$ is the scale of inflation, then its global symmetry will be broken after inflation and the subsequent phase transition will populate the universe with bubbles of different vacuum expectation values, forming a network of domain walls. These bubbles are expected to collapse to form structures called ``mini-clusters'' which can be the source of cosmological structure formation and/or black holes which may be the origin of the super massive black holes in the centers of galaxies \cite{Hogan:1988mp}.

In the early universe, potentials with many minima are often considered to be candidates for inflationary models (see \cite{Martin:2013tda} for an exhaustive review). In the context of Type IIB string theories, the low energy effective theory can often be described by a potential landscape of many different minima, arising from the different choice of fluxes used for its compactification to 4 dimensions \cite{Feng:2000if}. In \cite{Wainwright:2014pta}, collisions between pairs of bubbles are studied in 1+1D, and the observables resulting from such a ``multiverse'' scenario are quantified, such that possible models may be constrained by observables. However, in the early universe, many such bubbles may have formed and collided simultaneously, leading to more random and asymmetric configurations. This paper represents a first step towards understanding these more complex interactions. In this context, a single asymmetric bubble collapse can act as a simple model for the collapsing shapes formed when multiple spherical bubbles collide randomly and simultaneously. We will study such multiple bubble collisions in a future work. 

The paper is organised as follows:
\begin{itemize}
\item In section \ref{sect:formcode}, we briefly describe the methodology and formalism used in the 3+1 simulations. The 1D formalism is given in Appendix \ref{sect:appx}.
\item In section \ref{sect:1Dout}, we describe the spherical symmetric case, using both a 1+1D code and the full 3+1D simulations with GRChombo.
\item In section \ref{sect:3Dout}, we describe collapse in two different asymmetric cases. 
\item In section \ref{sect:discuss}, we discuss the results and suggest areas for further work.
\end{itemize}

\section{Methodology}
\label{sect:formcode}

In this section we briefly describe the methodology and formalism used in the 3+1 simulations. The 1D formalism used is given in Appendix \ref{sect:appx}.

\subsection{Gauge conditions}

\begin{figure}
\begin{center}
\includegraphics[width=10cm]{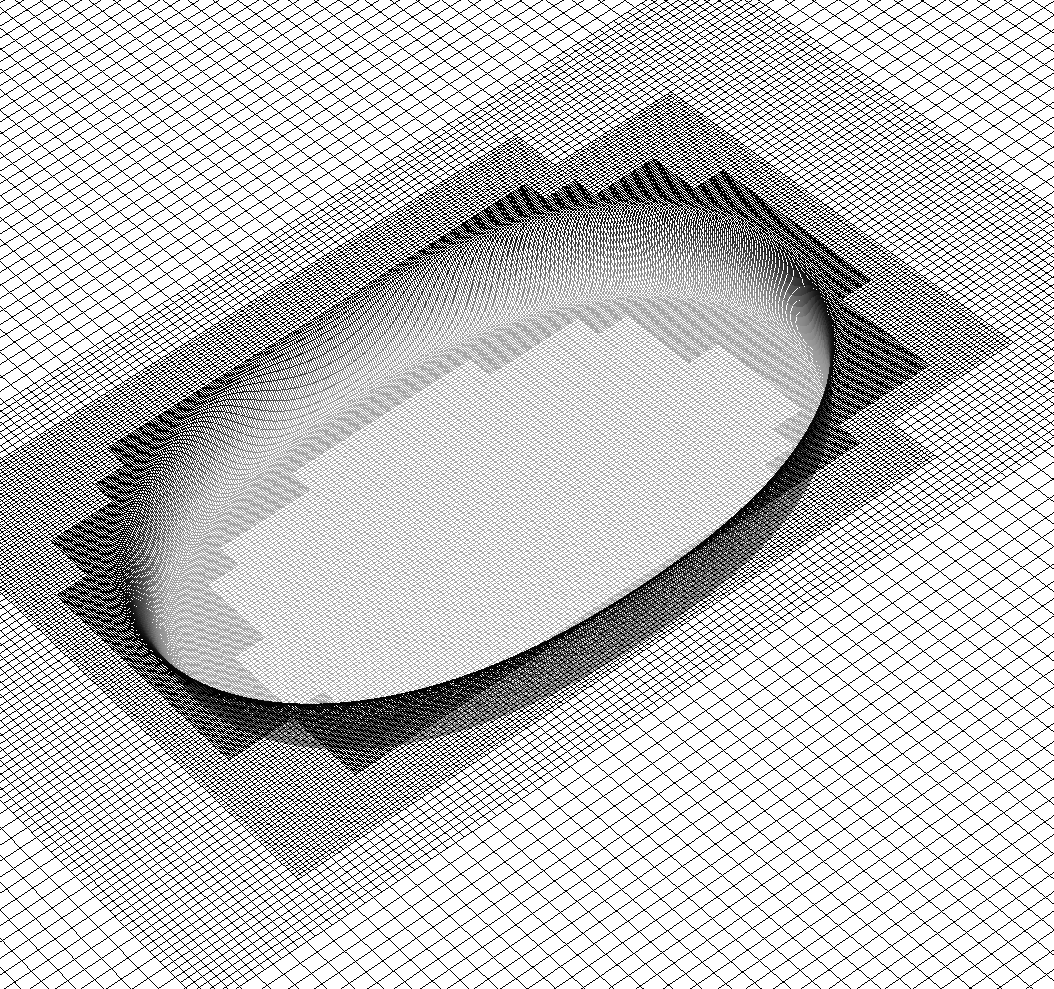}
\caption{The mesh for a scalar field bubble in GRChombo, which is adapted to give greatest curvature at the steep walls of the bubble. The image shows a slice through the x-z plane with the elevation corresponding to the value of the field $\phi$ at that point.
\label{fig-mesh}}
\end{center}
\end{figure}

A full description of the GRChombo code can be found in \cite{Clough:2015sqa}. Briefly, the code implements a fourth-order accurate BSSN formulation of the 3+1D decomposition of the Einstein equation, with non-trivial ``many-boxes-in-many-boxes'' mesh hierarchies and massive parallelism through the Message Passing Interface (MPI). An illustration of the adaptive mesh responding to the curvature in an axisymmetric bubble is shown in Figure \ref{fig-mesh}.

In this paper, a modification of the moving puncture gauge condition, see \cite{Campanelli:2005dd, Baker:2005vv}, was used in all the evolutions. It was found that having steep walled bubbles, and adding in a potential term, made the system of equations significantly more challenging to evolve stably near the critical point. After some experimentation with different gauges, the most suitable gauge was found to be given by the Gamma driver condition for the shift,
\begin{eqnarray}
\partial_t \beta^i &=& B^i\, \label{eqn:betadriver}\\
\partial_t B^i &=& \mu_{\beta_1}\alpha^{\mu_{\beta_2}}\partial_t \hat\Gamma^i-\eta B^i\ , \label{eqn:gammadriver}
\end{eqnarray}
where $\mu_{\beta_1}=1$, $\mu_{\beta_2}=0$ and $\eta$ is of order $1/2M$, where $M$ is the ADM mass of the initial spacetime, and the following condition for the lapse,
\begin{equation}
\partial_t \alpha = -\mu_{\alpha_1}\alpha^{\mu_{\alpha_2}}K + \mu_{\alpha_3}\beta^i\partial_i \alpha, \label{eqn:alphadriver}\,
\end{equation}
with $\mu_{\alpha_1}$ of order 1, but set as described below, $\mu_{\alpha_2}=2$ and $\mu_{\alpha_3}=1$. 

Note that this lapse condition would be the harmonic gauge if coupled with a zero shift vector. It was found that using $\alpha^2$ as opposed to simply $\alpha$ (as in the standard puncture gauge) kept the lapse around ${\cal O}(1)$ during the final stages of the collapse and thereby increased stability.

As noted, it was non-trivial to find a stable gauge for evolutions near the critical point, possibly due to the additional non-linearity introduced by the self interaction potential, and a certain amount of trial and error was required. The key requirements for stability are listed below.

\begin{itemize}
\item The coefficients in the lapse condition were chosen so that the lapse remained between values of 0.1 and 1.0 until the field settled into a black hole or dispersed, as freezing of the coordinates or large forward steps resulted in instabilities developing. In the late stages of collapse high scalar field gradients resulted in the lapse being driven near zero at the centre of the grid, whereas it needed to be free to oscillate further in order to prevent slice-stretching. The $\mu_{\alpha_1}$ parameter for the lapse condition was therefore chosen to keep the lapse of order one whilst the solution was continuing to evolve. This was found to be important for stability close to the critical point.
\item A non-zero, evolving shift was required in supercritical cases for the evolution to remain stable. The $\eta$ coefficient of the shift was of order $1/2M$ where $M$ is the ADM mass of the initial spacetime, but the stability of simulations was not particularly sensitive to its exact value. 
\item Some level of damping of high frequency numerical noise, such as by Kreiss-Oliger dissipation \cite{TUS:TUS1547}, was needed, but again the exact coefficient was not crucial to stability.
\end{itemize}

\subsection{Initial data}

In the spherically symmetric case the initial conditions were derived from a numerical Mathematica solution in the areal polar gauge (as described in the Appendix). 

In asymmetric cases, we chose for the initial conditions a moment of time symmetry such that $K_{ij} = 0$, and a conformally flat metric. The remaining degree of freedom, the conformal factor $\chi$, was solved for using a relaxation of the Hamiltonian constraint $\mathcal{H}$ over some chosen relaxation time $\tau$, that is,
\begin{eqnarray}
\partial_{\tau} \chi = \mathcal{H} .
\end{eqnarray}
Note that in GRChombo $\chi$ is defined by $\gamma_{ij}=\chi^{-2}\,\tilde\gamma_{ij}$ such that $\chi = \left(\det\gamma_{ij}\right)^{-1/6}$.

\subsection{Resolution and convergence}

The coarsest level of refinement was $dx=M$, with $M$ an arbitrary mass scale in the simulation (we work in geometric units in which $G=c=1$). The physical domain was $(128M)^3$ and regridding was triggered by the change in $\phi$ or $\chi$ across the cell exceeding a certain empirically determined threshold (see \cite{Clough:2015sqa} for details). The maximum number of regriddings was 8, 9 or 10 depending on how close the simulation was to the critical point. In supercritical evolutions, the number of grid points across the event horizon needed to be, at minimum, around 20 in order for the horizon to be well resolved. A Courant factor of 0.25 was used in the fourth order Runge-Kutta update.

It was found that increasing the number of levels of refinement, and reducing the threshold at which regridding occurred, did not change the mass of the black hole formed, so that the results had converged at the levels used. The black hole masses were measured from the area of the apparent horizon, once the black hole had settled into a spherically symmetric configuration (for asymmetric initial conditions) and the majority of the scalar field radiation had either dispersed or fallen into the black hole. 

The typical run time for a full collapse into a black hole was of the order of 24-36 hours when run on 256 cores. GRChombo exhibits strong scaling up to around 1,000 cores and so this can be reduced if more resources are available. Closer to the critical point the simulations were more challenging and above 11 levels (10 regriddings above the coarsest level) the simulations because impractically slow and memory intensive with the resources we had available. However, there was in principle no barrier to increasing the levels of refinement further to study smaller mass black holes.

\section{Spherical Symmetry}
\label{sect:1Dout}

\subsection{Spherical symmetry in 1+1D}

In preparation for the full 3+1D GR simulations, we simulated bubble collapses in spherical symmetry with a separate 1+1D numerical code, with an initial configuration as in equation \ref{eqn:phiinit}, subject to the potential in equation \ref{eqn:potential} with $s=1$, $\phi_0=0.01$, and $\zeta = 10,000$. 

\begin{figure}
\begin{center}
\includegraphics[width=10cm]{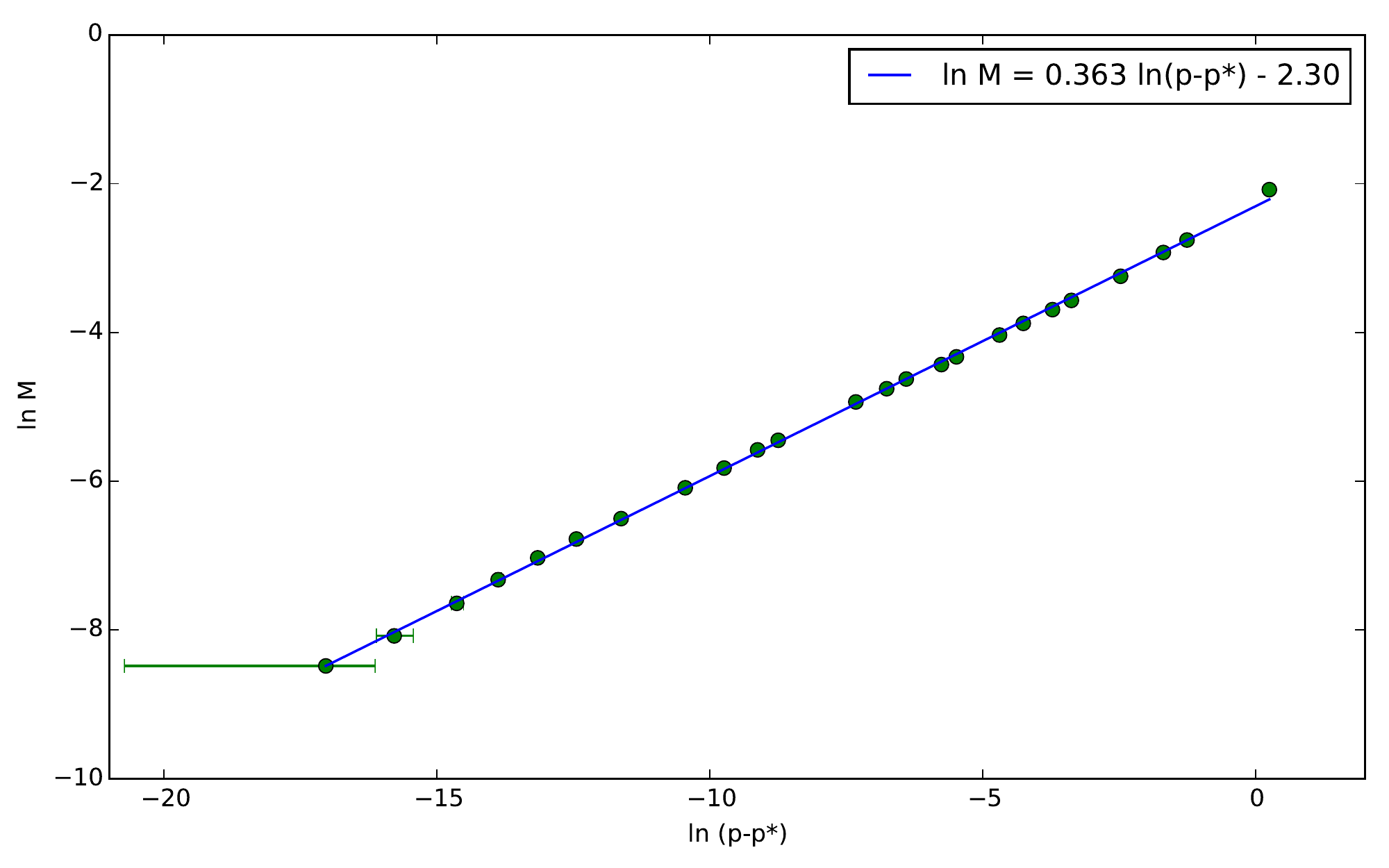}
\caption{Results from the simulations of a 1+1D spherically symmetric code. The critical index is $\gamma=0.363$, which is consistent with the value obtained in the full 3+1D simulations presented elsewhere.}
\label{fig-1Dcode}
\end{center}
\end{figure}

The value of the initial bubble radius $R_0$ was chosen as the critical parameter, and was increased from subcritical to supercritical. The masses of the resulting black holes were recorded.  The results are given in figure \ref{fig-1Dcode}. The critical index from these simulations is $\gamma=0.363$, consistent over a range of $\ln (R-R_*)$ from $0$ to $-15$. There is an uncertainty of $+0.008$ and $-0.028$, which arises from considering the uncertainty in the critical point, as bounded by the smallest radius simulated for which a black hole formed, and the largest for which it did not. The quoted critical index is based on the critical value for which the residuals in the best fit line were minimised. There are hints of a periodic oscillation of the black hole masses as predicted in \cite{Gundlach:1996eg} but we did not pursue this further.

A technical difficulty encountered during the simulations was the rapid Lorentz contraction of the bubble wall during the collapse, which necessitated very high resolution meshes. Since most of the energy of the bubble wall is concentrated along a small region of the simulation domain, this posed a challenge even in 1+1D. We used a variable mesh 1+1D code (as opposed to general adaptive mesh) to push as close to the critical point as we could within reasonable expenditure of computational resources. This portended the difficulties we would face when we moved to full 3+1D simulations. 

\subsection{Spherical symmetry in 3+1D}

Initially, we repeated the $\zeta = 10,000$ tests undertaken with the 1+1D code in order to confirm that the 3+1D code gave consistent results.

We looked for a consistent scaling relation for black hole masses in supercritical collapses, and for evidence of scale echoing. For the latter, we observed the values of scale invariant quantities like $\chi$, $K$ and $\rho$ at the centre of the bubble in a slightly subcritical evolution, and how they evolved in proper time before the critical accumulation point was reached (the point at which the field began to disperse). We also looked at radial profiles of $dm/dr = 4\pi r^2 \rho$ at and around the critical time.

\begin{figure}
\begin{center}
\includegraphics[width=10cm]{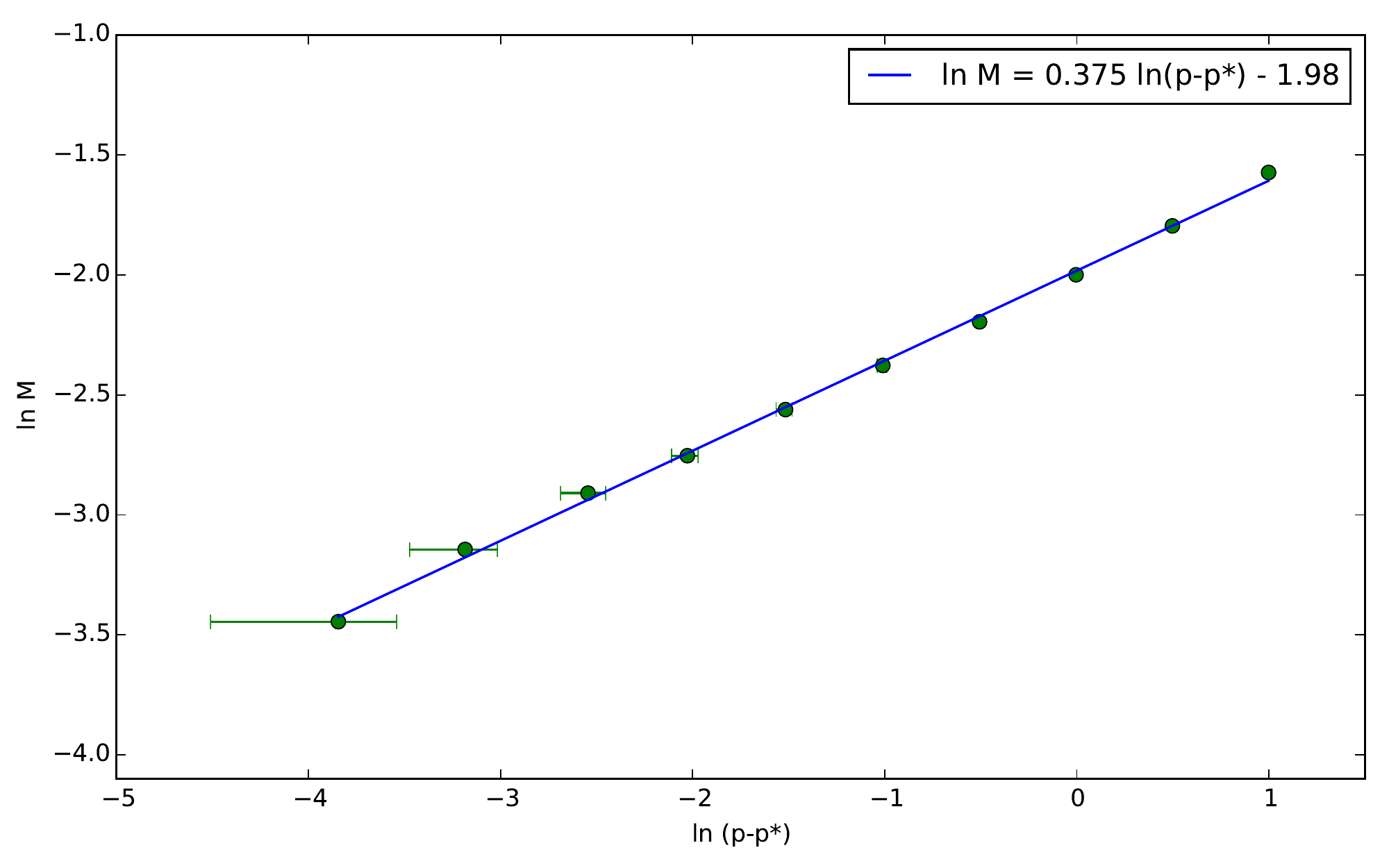}
\caption{Plot of ln$(M_{BH})$ vs ln$(R-R^*)$ for the symmetric 3+1D bubble in a potential with $\zeta =$ 10,000. The horizontal error bars represent the possible error in the plotted values due to the uncertainty in the critical value of $p^*$. The best fit line is based on the critical value for which the residuals of the fit were minimised.
\label{fig-scale1}}
\end{center}
\end{figure}

We found that, as was expected, spherical bubbles subject to a double well potential showed the same critical phenomena as in the massless case and were consistent with our 1+1D simulations. Figure \ref{fig-scale1} shows a plot of the scaling relation between ln$(M_{BH})$ and ln$(R-R^*)$ which was obtained, with the best fit line giving a value for the critical exponent of $\gamma = 0.375$. There is an uncertainty of $+0.020$ and $-0.037$ which arises from considering the uncertainty in the critical point, as bounded by the smallest radius simulated for which a black hole formed, and the largest for which it did not. As in the 1D case, the quoted critical index is based on the critical value for which the residuals in the best fit line were minimised. 

\begin{figure}
\begin{center}
\includegraphics[width=15cm]{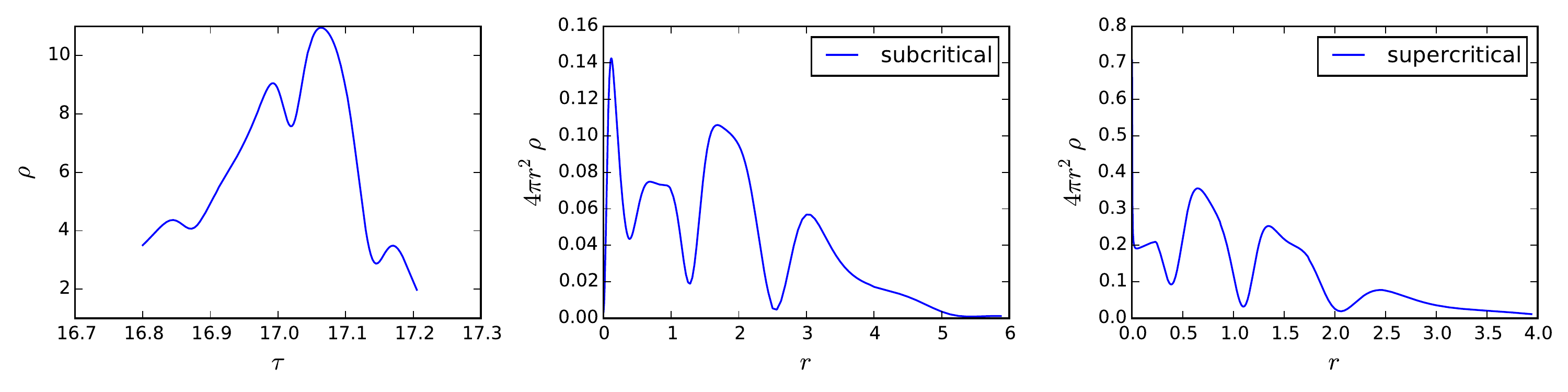}
\caption{Plots of echoing for the symmetric bubble in a potential with $\zeta =$ 10,000. The first shows the evolution of the energy density $\rho$ against proper time $\tau$. If one applies a fit to $\tau^*$ and $\Delta$ using the relation in the echoing equation with respect to the extremal values in this plot one finds a value of $\Delta$ of approximately 0.7, which is inconsistent with the massless case value of 3.4 which is expected. The second and third plots show radial profiles of $dm/dr = 4\pi r^2 \rho$ in the marginally subcritical and supercritical cases simulated, and whilst there are hints of echoing, they are not conclusive.
\label{fig-echo1}}
\end{center}
\end{figure}

Figure \ref{fig-echo1} shows the echoing plots that were produced in near critical evolutions. As can be seen, there is some evidence for echoing, but insufficient to be conclusive. If one applies a fit to $\tau^*$ and $\Delta$ using the relation in Equation \ref{eqn:echo}, one finds a value of $\Delta$ of approximately 0.7, which is inconsistent with the massless case value of 3.4 which is expected. It is likely that we are still too far from the critical point to observe true echoing. It is also possible that, in the case of the radial profiles, the chosen coordinates are not well adapted to the echoing, as the value of the shift is non-zero, meaning that the coordinate value of $r$ will represent different physical points as the simulation progresses. 

\section{Beyond Spherical Symmetry - 3+1D simulations}
\label{sect:3Dout}
\subsection{Radial perturbations of a spherical bubble - axisymmetric bubbles}

\begin{figure}
\begin{center}
\includegraphics[width=10cm]{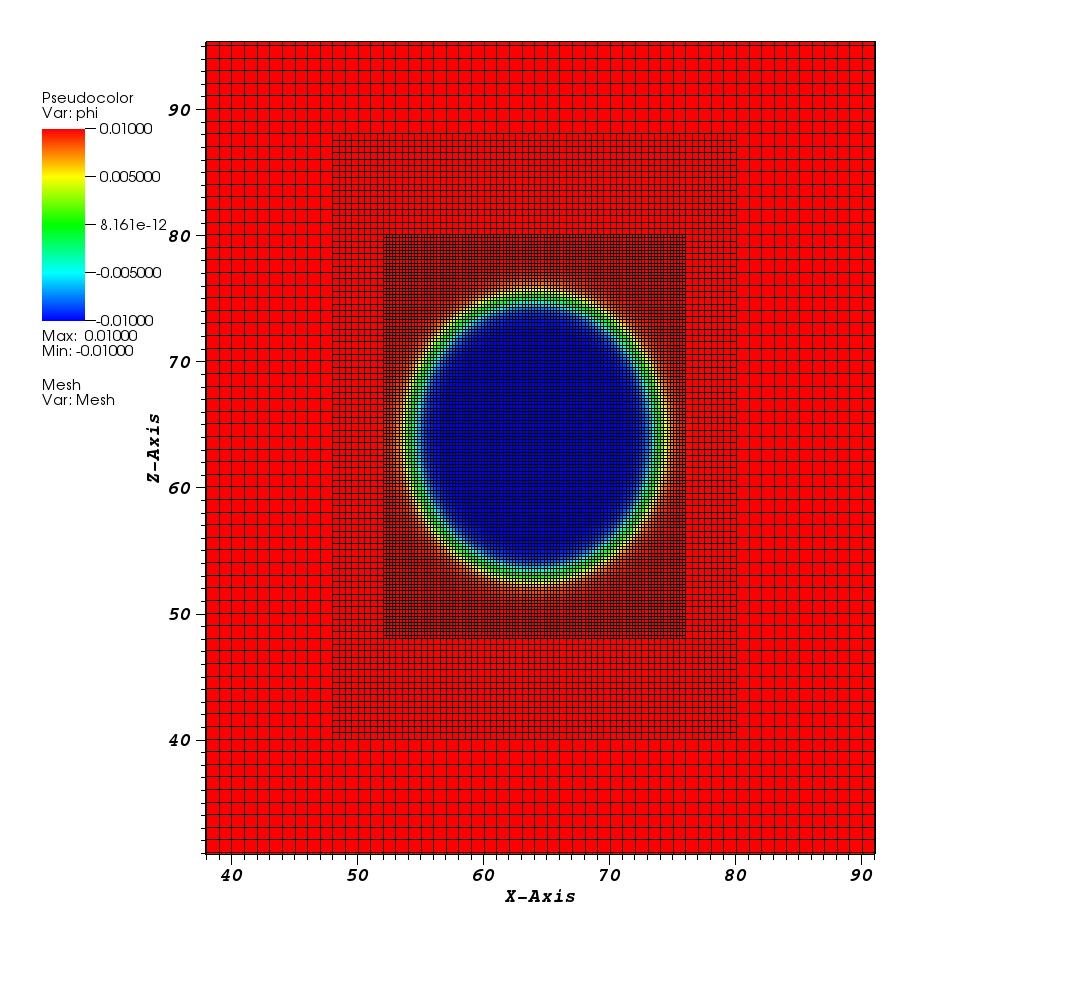}
\caption{Initial configurations for the axisymmetric radially perturbed bubble case, showing the adaptive mesh. The image shows a slice through the $x-z$ plane with the colour corresponding to the value of $\phi$ at that point per the legend.
\label{fig-bubaxi}}
\end{center}
\end{figure}

\begin{figure}
\begin{center}
\includegraphics[width=10cm]{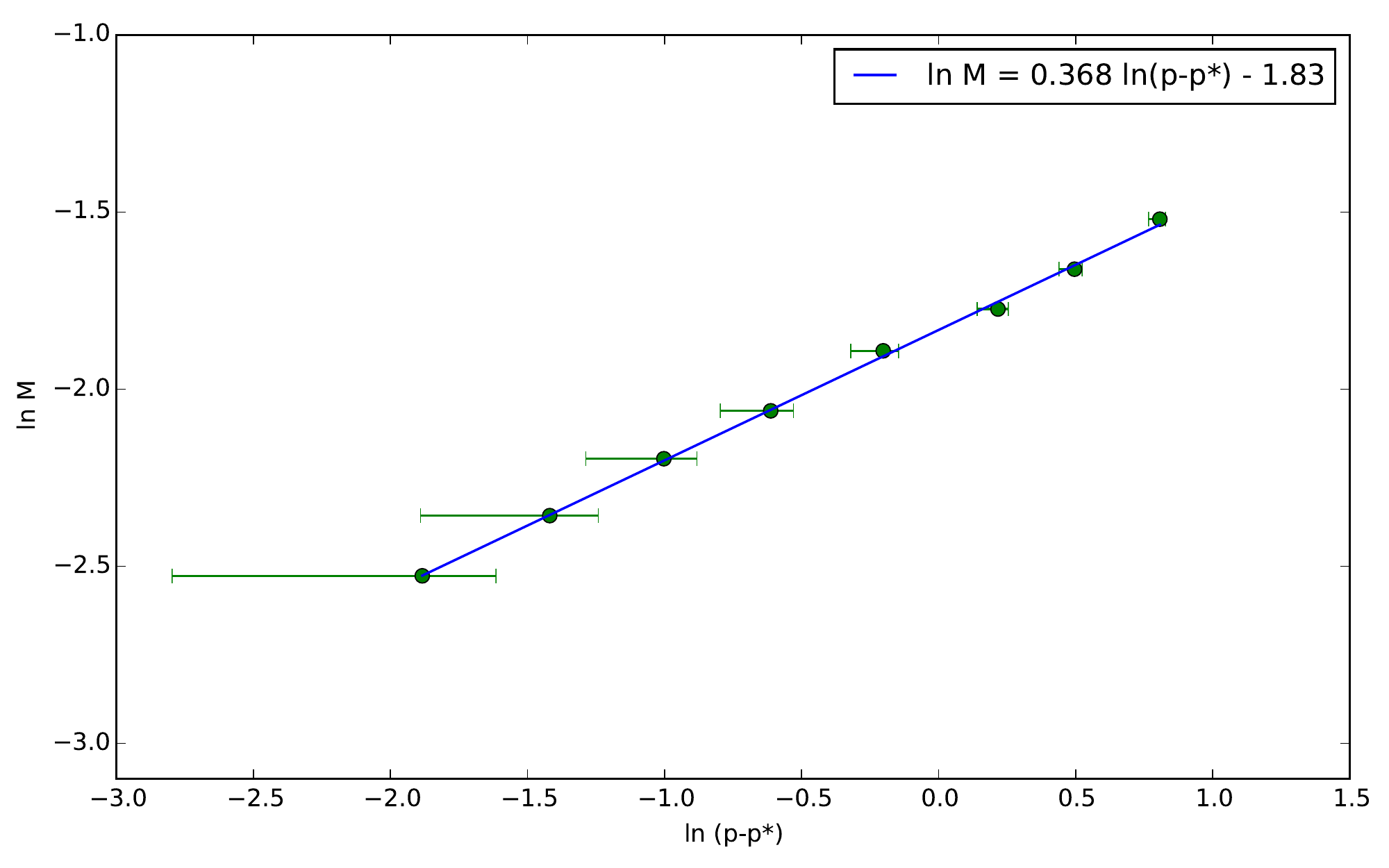}
\caption{Plot of ln$(M_{BH})$ vs ln$(R-R^*)$ for the axisymmetric, radially perturbed bubble in a potential with $\zeta =$ 5,000. The horizontal error bars represent the possible error in the plotted values due to the uncertainty in the critical value of $p^*$. The best fit line is based on the critical value for which the residuals of the fit were minimised.
\label{fig-scale2}}
\end{center}
\end{figure}

We set up bubbles in a $s=1$, $\phi_0=0.01$, and $\zeta = 5,000$ potential for which the radius of the bubble wall was perturbed by the $Y^1_0$ spherical harmonic, such that the initial configuration for the field is (see Figure \ref{fig-bubaxi}))
\begin{eqnarray}
R_{asym} = \left( 1+\epsilon | Re(Y^1_0) |^2 \right) R_0 ,
\end{eqnarray}
\begin{eqnarray}
\phi = \phi_0 \tanh \left[k(R-R_{asym})\right], \label{eqn:phiinit2}
\end{eqnarray}
with k as in equation \ref{eqn:littlek}, which is initially static, i.e. 
\begin{eqnarray}
\Pi = 0
\end{eqnarray}
where
\begin{eqnarray}
\Pi = \frac{1}{\alpha}( \partial_t \phi - \beta^i \partial_i \phi ).
\end{eqnarray}
We considered the case in which $\epsilon = 0.5$, and again, we varied the initial radius of the bubbles $R_0$ until we had bounded the critical point. We investigated the scaling relations in the black hole masses which resulted in supercritical evolutions, and scale echoing in evolutions above and below the critical point. 
We found evidence for critical phenomena consistent with that in the massless and spherically symmetric cases. Figure \ref{fig-scale2} shows a plot of the scaling relation between ln$(M_{BH})$ and ln$(R-R^*)$ which was obtained, with the best fit line giving a value for the critical exponent of $\gamma = 0.368$, using the method described in spherical symmetry above. This value is consistent with the spherically symmetric case. The uncertainty in this case was $+0.035$ and $-0.084$.

\begin{figure}
\begin{center}
\includegraphics[width=15cm]{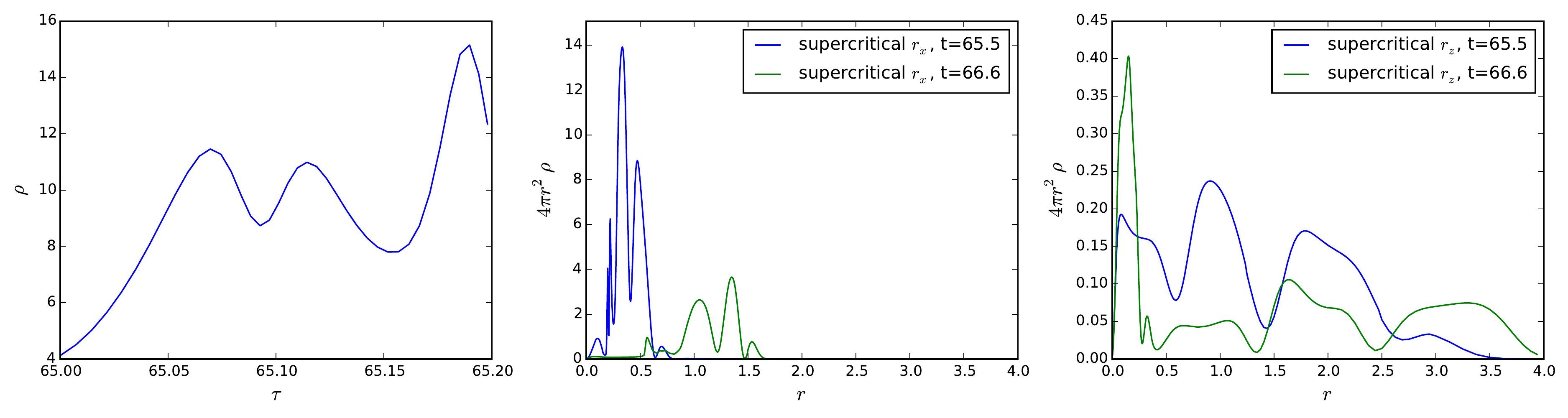}
\caption{Plots of echoing for the axisymmetric, radially perturbed bubble in a potential with $\zeta =$ 5,000. The first graph shows the evolution of the energy density $\rho$ against proper time $\tau$. The peaks are not consistent with the relation in the echoing equation since the proper time between the peaks appears to increase as the critical time is approached. The second plot shows radial profiles of $4\pi r^2 \rho$ along the z axis in a subcritical case, at two times. The third shows the same profiles in the x axis, perpendicular to the axis of symmetry.
\label{fig-echo2}}
\end{center}
\end{figure}

Figure \ref{fig-echo2} shows the echoing plots that were produced. Again, there is some evidence for echoing, but insufficient to be conclusive, for the reasons discussed above.

\begin{figure}
\begin{center}
\includegraphics[width=10cm]{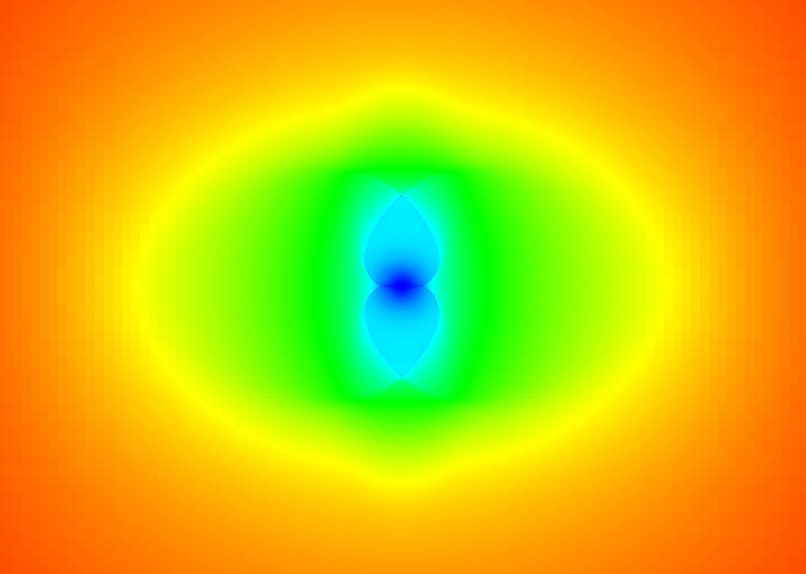}
\caption{We find that ``shock waves'' (blue regions transitioning into green) develop in some parameters in the axisymmetric case close to the critical point, particularly in $\alpha$, the lapse parameter, which is shown here, with very steep gradients and asymmetric configurations.
\label{fig-shock}}
\end{center}
\end{figure}

Although we saw no evidence in the simulations for other modes, it was clear that the behaviour was becoming more strongly asymmetric as the critical radius was approached, and the simulations became extremely challenging and difficult to evolve as the critical point was neared. We saw in particular that ``shock waves'' developed in some parameters, particularly in the lapse parameter, with extremely steep gradients and asymmetric configurations, as shown in figure \ref{fig-shock}. To enable us to probe this further, we are considering amending the gauge conditions further to ``smooth'' the lapse, and perhaps to implement some of the shock avoidance techniques applied in \cite{Figueras:2015hkb}.

\subsection{Amplitude perturbations of a spherical bubble - asymmetric bubbles}

\begin{figure}
\begin{center}
\includegraphics[width=10cm]{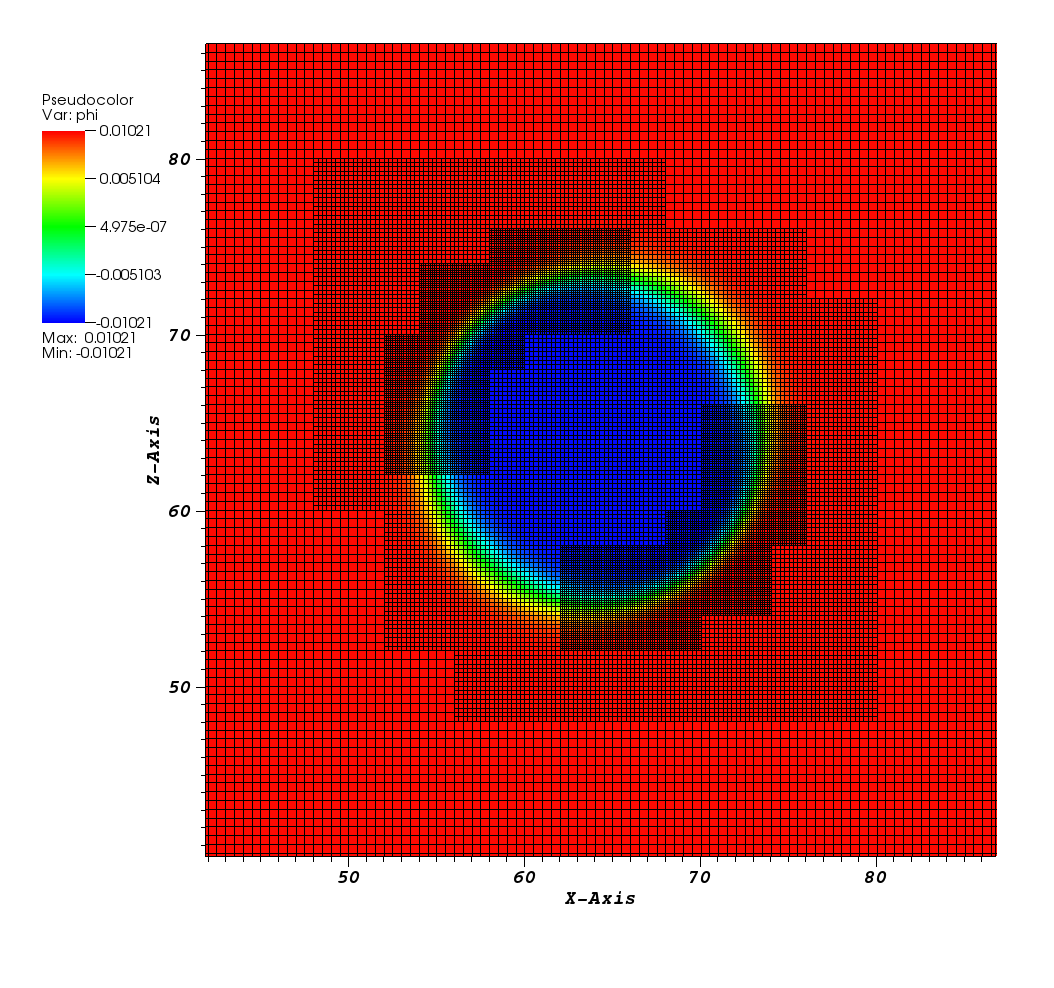}
\caption{Initial configurations for the asymmetric amplitude perturbed bubble case, showing the adaptive mesh. The image shows a slice through the x-z plane with the colour corresponding to the value of $\phi$ at that point per the legend.
\label{fig-bubasym}}
\end{center}
\end{figure}

We set up bubbles in a $s=1$, $\phi_0=0.01$, and $\zeta = 5,000$  potential for which the amplitude of the bubble at the wall was perturbed by the $Y^2_1$ spherical harmonic, similarly to \cite{Healy:2013xia}, such that the initial configuration for the field is (see Figure \ref{fig-bubasym})
\begin{eqnarray}
\phi = \left(1+\epsilon Re(Y^2_1)  e^{\frac{R-R_0}{\sigma}}\right)\phi_0 \tanh\left[k(R-R_0)\right], \label{eqn:phiinit3}
\end{eqnarray}
with k as in equation \ref{eqn:littlek}, and $\sigma=0.5$, and
\begin{eqnarray}
\Pi = 0.
\end{eqnarray}
We considered the case $\epsilon = 1$, and again, we varied the initial radius of the bubbles $R_0$ until we had bounded the critical point above and below. We investigated the scaling relations in the black hole masses which resulted, and evidence for scale echoing. 

\begin{figure}
\begin{center}
\includegraphics[width=10cm]{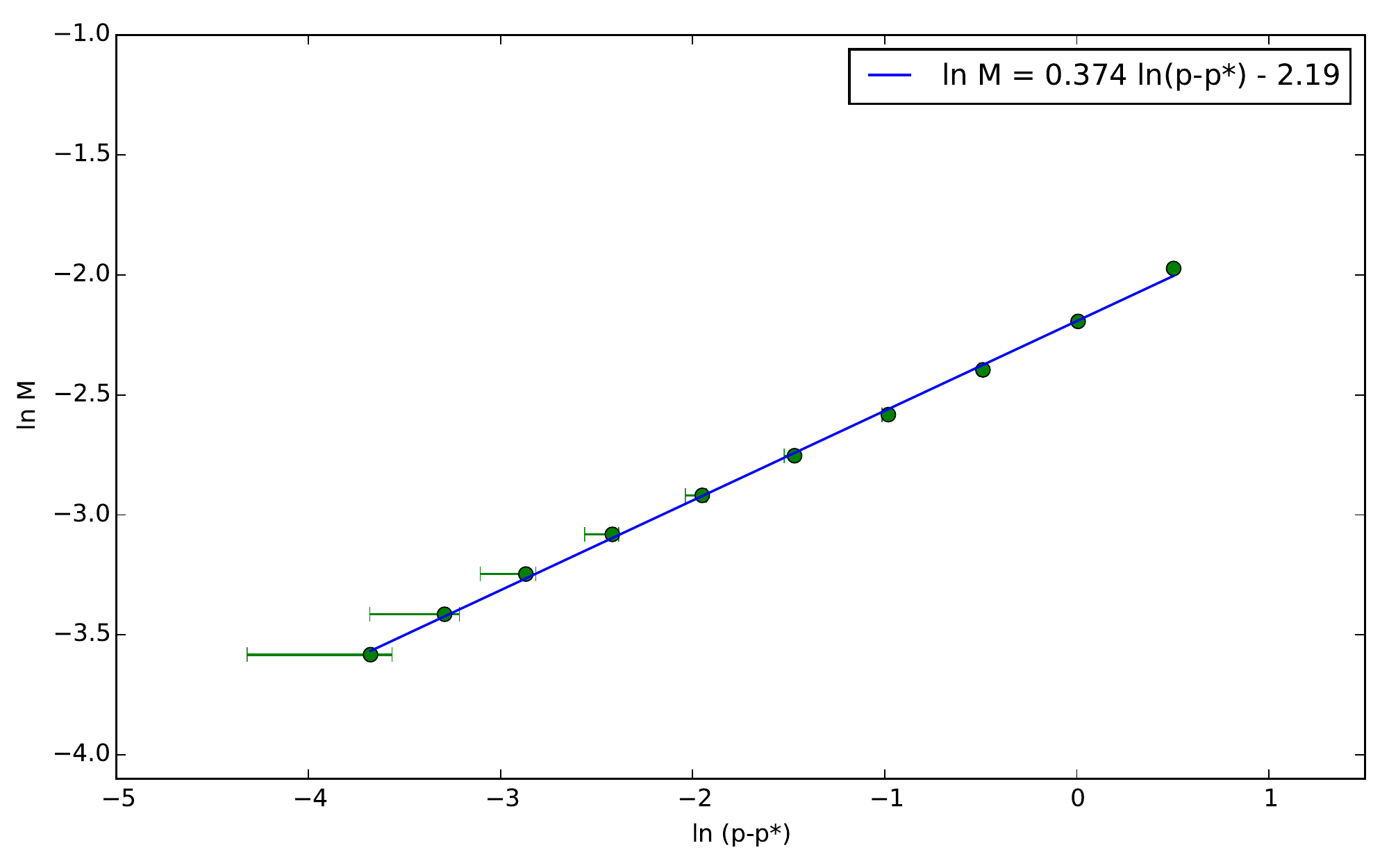}
\caption{Plot of ln$(M_{BH})$ vs ln$(R-R^*)$ for the asymmetric amplitude perturbed bubble in a potential with $\zeta =$ 5,000. The horizontal error bars represent the possible error in the plotted values due to the uncertainty in the critical value of $p^*$. The best fit line is based on the critical value for which the residuals of the fit were minimised.
\label{fig-scale3}}
\end{center}
\end{figure}

\begin{figure}
\begin{center}
\includegraphics[width=15cm]{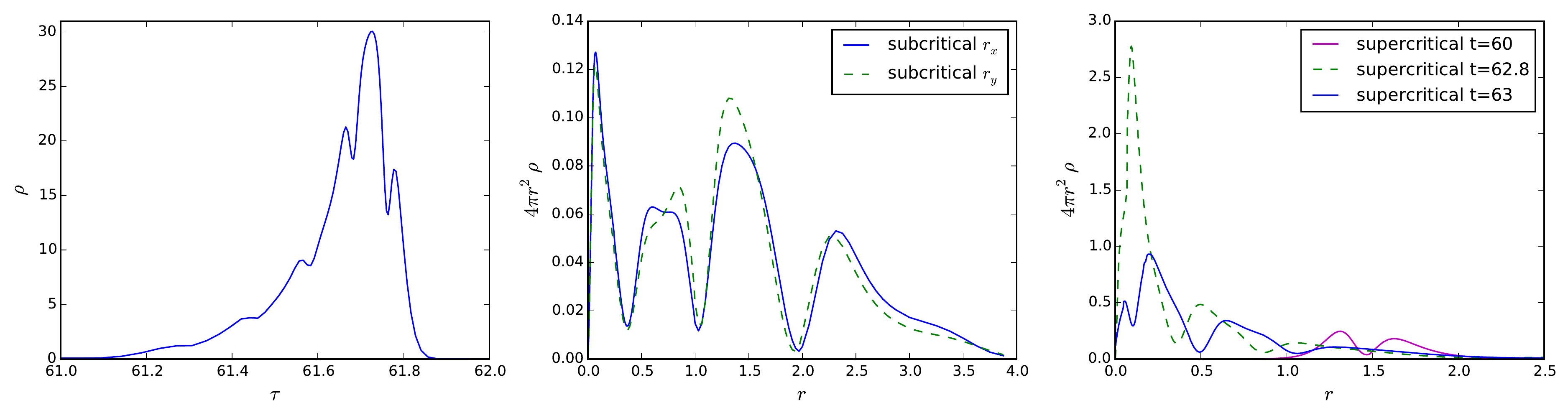}
\caption{Plots of echoing for the asymmetric amplitude perturbed bubble in a potential with $\zeta =$ 5,000. The first graph shows the evolution of the energy density $\rho$ against proper time $\tau$. If one applies a fit to $\tau^*$ and $\Delta$ using the relation in the echoing equation with respect to the extremal values in this plot one finds a value of $\Delta$ of approximately 0.7, which is inconsistent with the massless case value of 3.4 which is expected. The second plot shows radial profiles of $4\pi r^2 \rho$ along and perpendicular to the z axis in a subcritical case. The third shows the z axis profile in a supercritical case at several times close to the critical time. Whilst there are hints of echoing in these plots, they are not conclusive.
\label{fig-echo3}}
\end{center}
\end{figure}

We found that the same critical phenomena as in the symmetric and axisymmetric cases. Figure \ref{fig-scale3} shows a plot of the scaling relation between ln$(M_{BH})$ and ln$(R-R^*)$ which was obtained, with the best fit line which minimises the residuals giving a value for the critical exponent of $\gamma = 0.374$, with an uncertainty of $+0.009$ and $-0.043$.
Figure \ref{fig-echo3} shows the echoing plots that were produced in near critical evolutions. As above, there is some evidence for echoing, but insufficient to be conclusive. 

\section{Discussion}
\label{sect:discuss}

We have shown that in spherical symmetry, bubble collapse behaviour with a non-trivial self interaction shows, as expected, the same critical behaviour as in the massless case.

In the axisymmetric and asymmetric cases studied, in which the radius of the initial bubble and its amplitude, respectively, were perturbed, we see a scaling relation which is again consistent with the massless spherically symmetric case. This seems to imply that, at least for the cases we considered, even in the presence of fairly significant initial asymmetry in the configuration, the final state of the collapse is dependent on a single dominant mode. 

There are some hints of echoing, but insufficient to be conclusive. It is likely that we are still too far from the critical point to observe echoing. It is also possible that the chosen coordinates are not well adapted to the echoing.

One might expect that there should be a point at which the scaling relation will eventually break down. For example, if we set up an ellipsoidal shaped bubble,  then one can imagine that the bubble can collapse into two black holes if one of the semi-principal axes is much longer than the other two (i.e. a ``long sausage''). Nevertheless, these two black holes will eventually collide and merge to form a single black hole as the final state. It is an interesting question to ask whether this final state black hole will still follow the scaling relation. This is work currently in progress.

\section*{Acknowledgments}
We thank Adam Brown and Tom Giblin for useful conversations on a related project. We would also like to thank our GRChombo collaborators Pau Figueras, Hal Finkel, Markus Kunesch and Saran Tunyasuvunakool for their work on the code, and various useful insights, and Juha J{\"a}ykk{\"a} and James Briggs for their ongoing technical support.  Numerical simulations were performed on the COSMOS supercomputer, part of the DiRAC HPC, a facility which is funded by STFC and BIS. EAL acknowledges support from an STFC AGP grant ST/L000717/1. This work also used the ARCHER UK National Supercomputing Service (http://www.archer.ac.uk) for some simulations. PETSc  \cite{petsc-web-page} was used for the 1+1D code.
\appendix
\section{1+1D formalism}
\label{sect:appx}

We use a 1+1D constrained evolution to simulate spherically symmetric cases of bubble collapse. In such an evolution, the field values are evolved in time but the metric variables are solved for on each slice using the constraint equations, as detailed below. 

We use a metric adapted to spherical symmetry with shift $\beta =0$, following the convention of \cite{Alcubierrebook} (for which we follow the conventions and notation therein),
\begin{eqnarray}
ds^2 = - \alpha^2 dt^2 + A dr^2 + B r^2 d\Omega^2 .
\end{eqnarray}
The gauge is fixed by choosing $B$ to be one throughout the evolution, which describes the polar areal gauge. The single remaining degree of freedom in the metric, $A$, may then be solved for on each slice using the Hamiltonian Constraint, which simplifies to
\begin{eqnarray}
\partial_r A = A \left( \frac{(1-A)}{r}  + 8\pi r A \rho \right) , \label{eqn:AA}
\end{eqnarray}
with the boundary condition $A=1$ at $r=0$. The energy density $\rho$, in terms of the field variables (defined in equations \ref{eqn:field1} and \ref{eqn:field2} below), is
\begin{eqnarray}
\rho = \frac{1}{2A}(\Psi^2 + \Pi^2) + V .
\end{eqnarray}
Fixing $B$ to be one throughout the evolution sets the condition $\partial_t B = \partial_t \gamma_{\theta \theta} = 0$, which in turn sets a condition on the lapse at each timestep of
\begin{eqnarray}
\partial_r \alpha = \alpha \left( \frac{(A - 1)}{r} + \frac{\partial_r A}{2A} - \frac{8 \pi A}{r} \left[ S_{\theta \theta} - \frac{r^2}{2} (S - \rho) \right] \right) . 
\end{eqnarray}
This equation can be expressed in terms of the field variables (defined in equations \ref{eqn:field1} and \ref{eqn:field2} below) and eliminating the derivative of A as
\begin{eqnarray}
\partial_r \alpha = \alpha \left( \frac{(A - 1)}{2r} - 2 \pi r (\Pi^2 + \Psi^2 + 2AV) \right) . \label{eqn:alp}
\end{eqnarray}
Boundary conditions are chosen for alpha such that:
\begin{eqnarray}
\alpha(r=r_{MAX}) = 1 
\end{eqnarray}
which leads to a "collapse of the lapse" when a black hole is formed. Equations \ref{eqn:AA} and \ref{eqn:alp} are then ODEs for $A$ and $\alpha$ which can be integrated on each time slice, rather than evolved dynamically. 
 
The field variables include the field $\phi$, and the auxiliary fields
\begin{eqnarray}
\Psi = \partial_r \phi \label{eqn:field1}
\end{eqnarray}
and
\begin{eqnarray}
\Pi = \frac{A^{1/2}}{\alpha} \partial_t \phi . \label{eqn:field2}
\end{eqnarray}
The initial field configurations for the bubbles are chosen to be a moment of time symmetry such that
\begin{eqnarray}
\phi = \phi_0 \tanh\left(k(R-R_0)\right), \label{eqn:phiinit},
\end{eqnarray}
and
\begin{eqnarray}
\Pi = 0 .
\end{eqnarray}
with $k$ defining the steepness of the wall between the two vacua and being set by the form of the potential in \ref{eqn:potential} as in, for example, chapter 2 of \cite{EWeinBook} as
\begin{eqnarray}
k = \sqrt{\frac{s}{2 \phi_0^2 \zeta}} \label{eqn:littlek} .
\end{eqnarray}
The evolution equations for these quantities are derived from the Klein Gordon equation $\nabla_{\mu} \nabla^{\mu} \phi = \partial_{\phi} V$ and are
\begin{eqnarray}
\partial_t \phi = \frac{\alpha}{A^{1/2}} \Pi ,
\end{eqnarray}
\begin{eqnarray}
\partial_t \Psi = \partial_r \left(\frac{\alpha}{A^{1/2}} \Pi\right) ,
\end{eqnarray}
and
\begin{eqnarray}
\partial_t \Pi = \frac{1}{r^2} \partial_r \left(\frac{\alpha r^2}{A^{1/2}} \Psi\right) - \alpha A^{1/2} \partial_\phi V .
\end{eqnarray}
As a diagnostic, the mass function
\begin{eqnarray}
m(r) = \frac{r}{2} \left(1-\frac{1}{A}\right) ,
\end{eqnarray}
gives the mass contained within that areal radius, and an event horizon is indicated by
\begin{eqnarray}
m(r) = \frac{r}{2} ,
\end{eqnarray}
at any point, with $m(r)$ then giving the mass of the black hole which has been formed.

This set of equations is discretized on a 1D mesh linearly in $r$ and solved using finite differencing. The time-stepping equations are solved using a standard RK4 solver, while the spatial integrations of $A$ and $\alpha$ in between time steps are solved using a RK2 solver. 

\bibliography{all.bib}
\bibliographystyle{utcaps}

\end{document}